\documentclass[12pt]{article}

\usepackage{epsf}
\usepackage[dvips]{graphicx}

\begin{document}

\begin{center}
{\bfseries Direct Photons in Ion Collisions at FAIR Energies}

\vskip 5mm

S.M. Kiselev\footnote{Talk at Baldin ISHEPP XVIII, Dubna, 25 - 30 September 2006}

\vskip 5mm

{\small
{\it
Institute for Theoretical and Experimental Physics, \\
Moscow, Russia; E-mail: Sergey.Kiselev@cern.ch
}
}

\end{center}

\vskip 5mm

\begin{center}
\begin{minipage}{150mm}
\centerline{\bf Abstract}
Estimations of prompt photon production at FAIR (Facility for Antiproton 
and Ion Research) energies using the extrapolation of existing data are presented.
About $10^{-4}$ prompt $\gamma$ with $p_{t}>$2 GeV/c per Au+Au
central event at 25 AGeV are expected. With the planed beam intensity
$10^{9}/s$, 1\% interaction rate and 10\% centrality, at CBM (Compressed Baryonic Matter) 
experiment one can expect prompt $\gamma$ rate ~100/s.

Predictions for direct photons by some generators (PYTHIA, UrQMD,
RQMD, HSD, HIJING) are analyzed. One of the main sources of
direct photons (due to meson scatterings $\pi\rho\rightarrow\pi\gamma, 
\pi\pi\rightarrow\rho\gamma$) is not implemented in the heavy-ion
generators. Corresponding cross-sections for this source have been 
prepared for implementation into the HSD code.  

Main experimental methods to study direct photons (subtraction method, momentum correlations 
method and internal conversion method) are shortly reviewed. High intensity beam, 
good tracking and good $e^{\pm}$ particle identification of the CBM detector favor
to measure direct photons by all the methods.
\end{minipage}
\end{center}

\vskip 10mm

\section{Introduction}
The FAIR~\cite{FAIR} accelerators will provide heavy ion beams up to Uranium 
at beam energies ranging from 2 - 45 AGeV (for Z/A=0.5) and up to 35 AGeV for
Z/A=0.4. The maximum proton beam energy is 90 GeV. The planed ion beam 
intensity is $10^9$ per second.

The CBM~\cite{CBM} detector will have good possibilities for
vertex reconstruction, tracking and identification of particles (hadrons, leptons
and photons). Though direct photons are of great interest for the research program
of the CBM experiment, a feasibility study has not been done yet.

Direct photons are photons not from particle decays. On the quark-gluon level three 
subprocesses dominate: Compton scattering $g q \rightarrow \gamma q$, 
annihilation $q \bar q \rightarrow \gamma g$ and bremsstrahlung emission
$q q(g) \rightarrow q q(g) \gamma$. Photons from initial hard NN collisions
are named prompt photons. They can be described by perturbative QCD (pQCD).
In the case of other limit - thermalized system of quarks and gluons,
quark-gluon plasma (QGP), these photons are named thermal photons from QGP.
Obviously, there is nonequelibrium stage in nuclear-nuclear collisions.

On the hadron level there are a lot of meson scattering channels:
$\pi \pi \rightarrow \rho \gamma, \pi \rho \rightarrow \pi \gamma,
\pi K \rightarrow K^* \gamma, K \rho \rightarrow K \gamma, ...$
First two channels give most contribution. If hadron system is thermalized
these photons are named thermal photons from hadron gas (HG). 

Here we would like to estimate for the CBM energy: prompt $\gamma$ contribution,
$\gamma$ contribution from hadron sources (decays and rescatterings) and
possibility to explore the state-of-the-art experimental methods.

\section{Prompt photons}
Existing $pp \rightarrow \gamma X$ date~\cite{Compelation} cover the
energy range $\sqrt{s}=20-1800$ GeV. Thus, CBM can fill the gap
$\sqrt{s}<14$ GeV.

\begin{figure}[h]
\centerline{
\includegraphics[width=50mm,height=50mm]{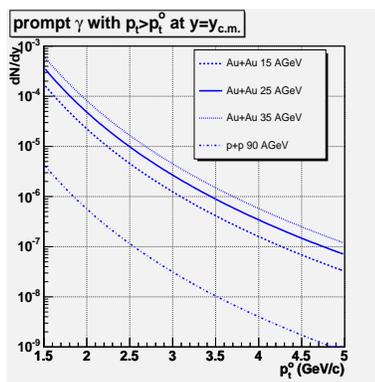}}
\caption{Yield of prompt photons at FAIR energies}
\label{fig:PROMPT}
\end{figure}

At transverse momentum $x_{T}=2p_{T}/\sqrt{s}>0.1$ cross sections can
be fitted in the central rapidity region by the formula
$Ed^{3}\sigma^{pp}/d^{3}p = 575(\sqrt{s})^{3.3}/p_{T}^{9.14}$ 
pb/GeV$^{2}$~\cite{Srivastava}. Using this fit one can estimate the
prompt $\gamma$ spectrum in nucleus-nucleus, A+B, collisions at impact
parameter b:
$Ed^{3}N^{AB}(b)/d^{3}p = Ed^{3}\sigma^{pp}/d^{3}p \cdot AB \cdot T_{AB}(b)$,
where the nuclear overlapping function is fefined as $T_{AB}(b)=N_{coll}(b)/\sigma^{pp}_{in}$,
where $N_{coll}(b)$ is the average number of binary NN collisions.
Fig.~\ref{fig:PROMPT} demonstrates yield of prompt photons at FAIR energies.   
For the central Au+Au events at 25 AGeV one can expect ~ $10^{-4}$
prompt photons with $p_{t}>2$ GeV/c. At beam intensity $10^{9}$ per second,
1\% interaction probability and 10\% of most central collisions
we can have 100 prompt $\gamma$ per second.

\begin{figure}[h]
\centerline{
\includegraphics[width=50mm,height=50mm]{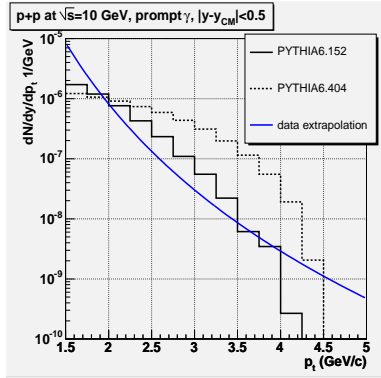}}
\caption{Spectra of prompt photons predicted by PYTHIA for p+p at $\sqrt{s}=10 GeV$}
\label{fig:PYTHIA-10}
\end{figure}

\begin{figure}[h]
\centerline{
\includegraphics[width=50mm,height=50mm]{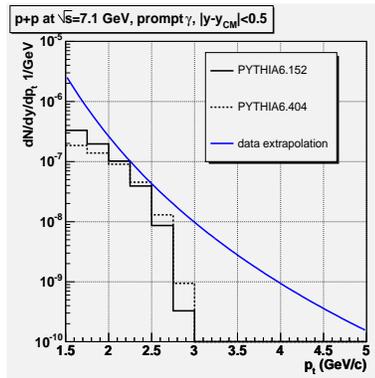}}
\caption{Spectra of prompt photons predicted by PYTHIA for p+p at $\sqrt{s}=7.1 GeV$}
\label{fig:PYTHIA-7}
\end{figure}

One can test the data extrapolation by PYTHIA~\cite{PYTHIA} simulations.
We have to switch on the subprocesses with photons: $g q \rightarrow \gamma q$,
$q \bar q \rightarrow \gamma g$ and $q \bar q \rightarrow \gamma\gamma$.
Fig.~\ref{fig:PYTHIA-10} shows spectra of prompt photons in pp collisions
at $\sqrt{s}=10$ GeV, lowest possible energy of the code. 
Though predictions of 6.152 and 6.404 versions differ at high $p_{t}$,
in the $p_{t}\approx2$ GeV/c range they agree with each other and with the
data extrapolation. To make simulation for the CBM energy one can change 
the lowest energy limit by $\sqrt{s}=7$ GeV. Predictions of the code are
presented in Fig.~\ref{fig:PYTHIA-7}.   
The PYTHIA values agree reasonably with the data extrapolation. For the prompt gamma
cross section we have the estimation: $\sigma \approx 2\cdot10^{-4}$ mb = 
 85\% $(g q \rightarrow \gamma q)$ + 15\% $(q \bar q \rightarrow \gamma g)$.
 
\section{photons from generators (UrQMD, HSD, RQMD, HIJING)}
The UrQMD transport code~\cite{UrQMD} was used in the paper~\cite{UrQMDdirectPhotons}
to evaluate direct photons for the Pb+Pb central collisions at CERN-SPS energy
158 AGeV. The processes $\pi \pi \rightarrow \rho \gamma, \pi \rho \rightarrow \pi \gamma$
were considered explicitly using cross sections given in~\cite{Kapusta}.
It was concluded that the rescattering $\pi \rho \rightarrow \pi \gamma$ and
the decay $\omega \rightarrow \pi \gamma$ processes are dominant in the range
1 GeV/c $< p_{t} <$ 3 GeV/c. In the sample of the central Au+Au events at 25 AGeV
generated by the CBM collaboration with the last version of UrQMD there are
14 photons per event. But all photons are from decays, mainly $a_{1} \rightarrow \pi \gamma$.
There are not photons from meson rescattereings at all. One of the UrQMD coauthor~\cite{Bleicher}
informed that users should ignore all processes with photons.

In the report~\cite{HSD} the HSD code was used to evaluate photons for the reaction
S+Au at 200 AGeV and compare its spectrum with the data. The code can generate
only decay photons. But in the sample of the central Au+Au events  at 25 AGeV
generated by the CBM collaboration with the free version HSD2.0 there are
no photons at all in the output file.

\begin{figure}[h]
\centerline{
\includegraphics[width=50mm,height=50mm]{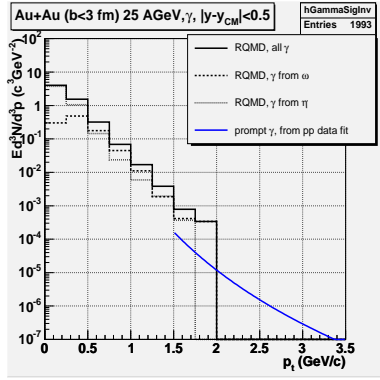}}
\caption{Spectra of photons predicted by RQMD for central Au+Au events at 25 AGeV}
\label{fig:RQMD}
\end{figure}

The RQMD code~\cite{RQMD} was also used to simulate $10^{3}$ central (b$<$3 fm) Au+Au 
events at 25 AGeV. The code generates only decay photons. In average there are 5.4 photons
per event, 66\% from $\eta'$ and 34\% from $\omega$ decays. Photon spectra are shown
in Fig.~\ref{fig:RQMD}  
One can see that at $p_{t} = 1.5 - 2$ GeV/c the contribution of photons from $\omega$
decays one order higher than from prompt photons estimated from the pp data.

The HIJING generator~\cite{HIJINGcode} takes into account as soft processes using
FRITIOF subroutines and hard ones calling the PYTHIA code. Though direct photons
production is included but at the CBM energy the code does not 
generate direct photons. One of the coauthor, X.N.Wang was informed about
this problem but up to now there is no a reply.  $10^{3}$ central (b$<$3 fm) Au+Au 
events at 25 AGeV have been generated using the code. There are 10.5 photons per
event, 60\% from $\eta'$, 36\% from $\omega$ and 4\% from $\Delta$ decays. 

Table~\ref{tab:CODES} summarize our tests to simulate direct photons using
different transport codes. Thus there is no a transport code generating direct photons.
\begin{table}[tbp]
\begin{center}
\begin{tabular}{|c|c|c|c|c|}
\hline
$\gamma$ origin &  UrQMD  &   HSD   &      RQMD     &   HIJING        \\
\hline
prompt          &    -    &    -    &       -       &   + BUT -       \\
\hline
decays          & + BUT - & + BUT ? &$\eta', \omega$&$\eta', \omega, \Delta, \phi, K^*$ \\ 
\hline
meson scatter.  & + BUT - &    -    &       -       &      -    \\
\hline
\end{tabular}
\end{center}
\caption{Photons from existing codes at the CBM energy (central Au+Au at 25 AGeV)}
\label{tab:CODES}
\end{table}

\begin{figure}[h]
\centerline{
\includegraphics[width=50mm,height=50mm]{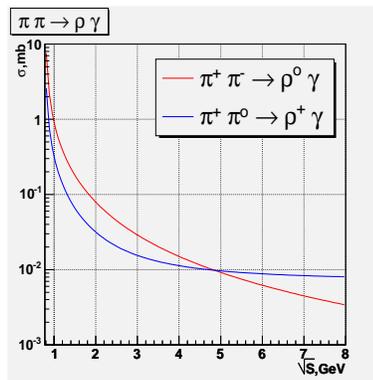}}
\caption{Cross sections for $\pi\pi\rightarrow\rho\gamma$}
\label{fig:PIPI}
\end{figure}

In agreement with the authors of the HSD generator we have prepared FORTRAN
subroutines with cross sections~\cite{Kapusta} for the processes $\pi\pi\rightarrow\rho\gamma$
and $\pi\rho\rightarrow\pi\gamma$, e.g. Fig.~\ref{fig:PIPI}. 
One can hope these subroutines will be soon implemented into the HSD code
and we shall have the first transport generator with direct photons due to meson
rescatterings.

\section{Experimental methods}
Extracting of direct photons in high-energy heavy-ion collisions
is very difficult an experimental task because of the large background 
from decay photons, mainly photons from $\pi^{0} \rightarrow \gamma\gamma$.
At low transverse momentum the task is more difficult due to larger
background.
Last years one can observe large progress in experimental techniques.
Direct photons have been revealed by three methods: subtraction method,
momentum correlation method and internal conversion method.

The first observation of direct photons in high-energy heavy-ion collisions have been presented by WA98 at 
SPS~\cite{subtractionWA98}. It was done by subtracting from the measured photons the background - photons from 
decays of reconstructed $\pi^{0}, \eta$ and other hadrons (with some assumption of its yield and spectrum). 
While peripheral Pb+Pb collisions exhibit no significant photon excess, the central reactions show a clear excess,
 ~20\%, of direct photons in the range of $p_{t}>2$ GeV/c. Using the same conventional method of decay photon 
subtraction, direct photons with large $p_{t}>4$ GeV/c have been observed at RHIC for Au+Au by PHENIX~\cite{subtractionPHENIX}. 
For the  $p_{t}$ range of 5 to 16 GeV/c the observed yield is consistent with a next-to-leading-order 
(NLO) pQCD calculation.

Then the WA98 collaboration has measured direct photons by the momentum correlations method~\cite{correlationWA98}. 
This HBT interferometry technique automatically eliminates the background from late hadron decays. The correlation 
strength parameter was used to determine the yield of direct photons at $p_{t}<$300 MeV/c. The measured yield exceeds 
theoretical expectations which attribute the dominant contribution in this $p_{t}$ region to the hadronic phase. 
To explain the large discrepancy is the real challenge for the theory. The correlation method needs larger statistics 
than the subtraction one. That is why results were obtained at the low $p_{t}$ range. For the other side the subtraction 
method failed in this range due to huge background.
 
Recently using a novel analysis technique based on low-mass $e^{+}e^{-}$ pairs with high $p_{t}$, PHENIX obtained interesting 
and intriguing results on direct photons. Exploiting the fact that any source of real photons emits also virtual photons, 
$\gamma^{*}$, with very low-mass, the low-mass $e^{+}e^{-}$ pair yield (after taking into account the contribution from Dalitz decays) 
is translated into a direct photon spectrum assuming  
$\gamma_{direct}/\gamma_{incl.}= \gamma^{*}_{direct}/\gamma^{*}_{incl.}$~\cite{internalConversionPHENIX}. 
Decay photons can mostly be eliminated by measuring the yield of $e^{+}e^{-}$ pairs in an invariant-mass region, 140 - 200 MeV, 
where pairs from the $\pi^{0}$ Dalitz decay are largely suppressed due to their limited phase space. Compared to the conventional 
measurement, this novel technique improves both the signal-to-background ratio and the energy resolution at intermediate $p_{t}$ 
where thermal production is expected to contribute substantially. Electrons in the central arms were identified by matching 
charged particle tracks to clusters in the electromagnetic calorimeter (ECAL) and to rings in the ring imaging Cherenkov (RICH) 
detector. The result is consistent with the conventional method, demonstrates smaller errors and shows a significant direct photon 
excess of about 10\% for 1 $<p_{t}<$ 5 GeV/c for central Au+Au collisions. The direct photons yield is significantly larger then 
the NLO pQCD expectation, but consistent with calculations when thermal photon emission is taken into account. However this 
theoretical interpretation is still debated. The CBM detector with its good tracking system, electron/positron identification 
by three detectors (a transition radiation detector (TRD), RICH and ECAL) and also photon identification by ECAL has very good 
possibility to use this method.

All three methods use ECAL, supplement each other and are power state-of-art instruments to study direct photons physics.

\section{Conclusions}
Estimations of prompt photon production at CBM energies using the extrapolation of existing 
data are presented. About $10^{-4}$ prompt $\gamma$ with $p_{t}>$2 GeV/c per Au+Au
central event at 25 AGeV are expected. With the planed beam intensity
$10^{9}/s$, 1\% interaction rate and 10\% centrality, one can expect prompt $\gamma$ rate ~100/s.

Predictions for direct photons by some generators (PYTHIA, UrQMD,
RQMD, HSD, HIJING) are analyzed. One of the main sources of
direct photons (due to meson scatterings $\pi\rho\rightarrow\pi\gamma, 
\pi\pi\rightarrow\rho\gamma$) is not implemented in the heavy-ion transport
generators. Corresponding cross-sections for this source have been 
prepared for implementation into the HSD code.  

Three experimental methods to study direct photons (subtraction method, momentum correlations 
method and internal conversion method) are shortly reviewed. They supplement each other and use 
ECAL as the main instrument. High intensity beam, good tracking and good $e^{\pm}$ particle 
identification of the CBM detector favor to measure direct photons by all the methods.

\section{Acknowledgments}
This  work  was partially supported by the Russian Foundation for Basic Research, 
grant number 06-08-01555 and Federal agency of Russia for atomic energy (Rosatom).

\section{Conclusions}

\end{document}